# Resolving the Intrinsic Bandgap and Edge Effect of BiI$_3$ Film Epitaxially Grown on Graphene

Dan Mu [a,1], Wei Zhou [b,1], Yundan Liu [a,*] Jin Li [a], Ming Yang [c], Jincheng Zhuang [c,*] Yi Du [c] and Jianxin Zhong [a,*]

[a] Hunan Key Laboratory of Micro-Nano Energy Materials and Devices, Xiangtan University, Xiangtan, Hunan 411105, PR China

[b] School of Electronic and Information Engineering, Changshu Institute of Technology, Changshu, Jiangsu 215500, China

[c] BUAA-UOW Joint Research Centre and School of Physics, Beihang University, Beijing 100191, China

[*] Authors to whom correspondence should be addressed: Y. Liu (liuyd@xtu.edu.cn); J.Zhuang (jincheng@buaa.edu.cn); J. Zhong (jxzhong@xtu.edu.cn)

[1] These authors contributed equally to this work

**Abstract:** Two-dimensional materials with layered structures, appropriate bandgap, and high carrier mobilities have gathered tremendous interests due to their potential applications in optoelectronic and photovoltaic devices. Here, we report the growth of BiI$_3$ thin film with controllable atomic thickness on graphene-terminated 6H-SiC(0001) substrate by molecular beam epitaxy (MBE) method. The growth kinetic processes and crystalline properties of the BiI$_3$ film are studied by scanning tunneling microscopy (STM). The scanning tunneling spectroscopy (STS) reveals a bandgap of 2.8 eV for monolayer BiI$_3$ with a weak dependence on film thickness for few-layer BiI$_3$, which greatly exceeds the previous reported values identified by macroscopic optical measurements. This discrepancy originates from the edge effect of BiI$_3$ that renders the bandgap downshift to 1.5 - 1.6 eV, as identified by the STS curves and the further confirmed by density functional theory (DFT) calculations. Our work provides a method to fabricate high-quality monolayer BiI$_3$ film and resolves its intrinsic bandgap as well as the edge effect on reduction of bandgap, benefitting not only to fundamental researches but also to nanoelectronic and optoelectronic applications.

# 1. Introduction

Two-dimensional (2D) semiconducting materials hold tremendous attention as a platform for energy storage systems, field-effect transistors, photonics and photovoltaic cells, inspired by their unique properties ascribed to the symmetry-broken after the thickness decreasing down to the single-layer limit [1-2]. As one of the promising candidates, Bismuth tri-iodide ($BiI_3$) exhibits a layered structure built from $BiI_6$ octahedrons with a rhombohedral crystal structure of $R\bar{3}$ symmetry, as shown in Figure 1b, where each unit layer possesses one Bi layer and two I layers with highly ionic Bi-I bonds and stacks along the out-of-plane direction under weak van der Waals interaction [3-4]. In the $BiI_6$ octahedrons, the central Bi atoms are symmetrically coordinated by six iodine atoms, giving rise to the lone pair of electrons on the $Bi^{3+}$ cation stereochemically inactive and consequently the semiconducting nature of $BiI_3$. Previously, $BiI_3$ thin film has been used for applications in X-ray detection, γ-ray detection, and pressure sensors because of its relatively wide bandgap, large mass density, and high effective atomic number [5-8]. Recently, it has been studied as a potential new type of metal-halide thin-film photovoltaic material due to the presence of antibonding states at the valence band edge, large absorption coefficient in the visible region of the solar spectrum, high electron mobility, and low carrier concentration [4,8-17]. For instance, $BiI_3$ is the prime precursor for preparing Bi-based perovskites as the adsorber layer in solar cells [18-20]. Compared to the $Pb^{2+}$ in conventional Pb-based perovskite materials, $Bi^{3+}$ in $BiI_3$ exhibits the similar electronic configuration but is nontoxic, leading to a green alternative for fabrication of stable, low cost, and high efficient solar cells [9-10,21-24]. Furthermore, the interface engineering of $BiI_3$ with other 2D materials to form the vertical heterostructures could improve its photovoltaic performance, as indicated by theoretical predictions and experimental results [12,25-26]. Nevertheless, there are several issues need to be illustrated before boosting potential applications of $BiI_3$. Firstly, the thin $BiI_3$ film with atomic thickness, which is expected to have intriguing electronic and optical properties distinguished from their bulk allotropes resulted from the quantum confinement effect, is a prerequisite for its integration in nanodevices. The experimental realization of monolayer or few-layer $BiI_3$ nanosheets, however, has not been achieved. Secondly, the literature reports on the bandgap of $BiI_3$ film studied by the macroscopic optical measurements are still under debate, with values in the range of 1.6-2.0 eV depending on different samples and different characterization techniques [3-4,6,8-10,14-15,27-28]. Resolving the intrinsic bandgap value of $BiI_3$ and the mechanism of its variation is essential for the band-structure engineering and practical promotion of this 2D material.

In this work, we report the van der Waals (vdW) epitaxy of $BiI_3$ thin films with the thickness ranged from monolayer to dozen layers on graphene surface on 6H-SiC(0001) by molecular beam epitaxy. The crystalline structure and electronic properties of the $BiI_3$ thin films have been revealed by STM and STS, and further confirmed by DFT calculations. From the STS curves, the intrinsic bandgap value of monolayer $BiI_3$ is found to be 2.8 eV with a weak dependence on film thickness, due to weak vdW interlayer interaction. Combined with the DFT calculation, the spatially resolved STS curves identify the edge effect on the electronic structure, which renders the bandgap downshift to a value close to the macroscopic optical results of 1.5eV - 1.6 eV.

## 2. Results and discussion

Figure 1a shows the STM image of the graphene-terminated 6H-SiC(0001) surface with a step height of 0.31 nm and an in-plane unit of 0.25 nm, corresponding to the thickness and lattice constant of single-layer graphene [29]. After deposition of $BiI_3$ for 10 min on the surface of graphene, the film with sub-monolayer coverage is identified in the STM image in Figure 1c. The inset of Figure 1c displays the line profile of the red dashed line in Figure 1c, inferring that the height of the island is about 0.71nm which is consistent with the thickness of single layer $BiI_3$ [11,17,26]. The high-resolution STM image of the deposited film, as shown in Figure 1d, exhibits the hexagonal arrangement with lattice constant of 0.78 nm, which agrees well with the predicted value of periodicity of monolayer $BiI_3$ but slightly larger than that of bulk materials (~0.75 nm) [11-12]. It should be noted that there are three top I atoms and two Bi atoms in the unit cell, as displayed in Figure 1d, which are probably detected by tunneling current during the scanning process. In order to figure out the detailed atomic structures of the deposited film and eliminate the effect of the density of states (DOS) in the STM images, we acquired STM images with atomic resolution in the same area by applying a sharp tip, as shown in Figure 1e. It can be seen that each protrusion in Figure 1d splits into three identical spots with the same brightness, indicating that the bright spots in Figure 1e are allied to the top I atoms and the protrusions in Figure 1d are contributed by the group effect of three nearest-neighboring top I atoms.

With the increment of deposition time, the whole graphene surface is covered by monolayer $BiI_3$ film, and the film grows then in three dimension, inducing rough multi-layer structures of islands, as shown in Figure 2a and Figure S1 in the Supporting Information, corresponding to the Stranski-Krastanov (SK) growth mode with the transition from 2D mode to 3D mode at one monolayer [30].

The STS curves in Figure 2b imply a large energy gap of 2.8 eV for monolayer $BiI_3$, and the gap value shows a weak dependence on film thickness. Figure 2c displays the bandgap as a function of layer numbers, where the gap value of the 14 layers (L) film reduces only 0.15 eV compared to that of monolayer sample. The previous calculations infer a slight gap deviation (~0.1-0.2 eV) between monolayer $BiI_3$ and its bulk allotrope obtained by various methods [11,26], due to the weak interlayer or interface interaction. The weak interlayer interaction is confirmed by the absence of Moiré pattern in the twisted adjacent $BiI_3$ layers in Figure 2d, as depicted by different orientations of the yellow rhombus standing for the unit cell of 5 L film (Figure 2e) and 6 L films (Figure 2f), respectively. The Moiré pattern is the geometric design of "superstructure" by superimposing one periodic pattern to another with relative lattice constant or spatial displacement. The Moiré pattern even exists in 2D materials with van der Waals interlayer interactions, such as bilayer graphene and silicene, providing homogeneous periodic potential with local variation for the whole system to form the superlattice-induced band structure in addition to the pristine constituent [31-34]. For the $BiI_3$ case, the large interlayer bond distance associated with weak interlayer interaction results in a weak relationship between its bandgap size and film thickness. Moreover, the atomic spatial arrangement is also identified between monolayer $BiI_3$ and underlying graphene without Moiré pattern in the STM images in Figure S2 (Supporting Information). All these results demonstrate that the monolayer $BiI_3$ is an "inert" material not only to itself but also to other 2D materials.

It should be noted that the bandgap of thick $BiI_3$ film derived from our STS curves is much larger than the previous reported values characterized by macroscopic optical measurements [3-4,6,8-10,14-15,27-28]. Thus, it is crucial to figure out the origin of this discrepancy prior to further investigations on this 2D material. The STS curves are applied to determine the local density of states of the material, while the results from optical characterizations are contributed by the integral outcome of the electronic structure of the whole sample. Therefore, any states, which may be originated from defects, strain, and edges, residing inside the gap between valence band maximum (VBM) and conduction band minimum (CBM) are expected to remarkably diminish the gap value in the optical measurements. In fact, there are plenty of monoiodine vacancies marked by red arrows in Figure 3a regardless layer numbers, which have been identified in scanning transmission electron microscopy in powder samples [35]. The STS curves collected on the pristine $BiI_3$ surface (labelled by point 1) and monoiodine vacancy position (labelled by point 2) in Figure 3b show almost the same behavior, demonstrating that the monoiodine vacancy has neglectable effect on the bandgap of $BiI_3$. This deduction is further supported by the line profile consisting of 20 spectra collected along the dashed arrow in Figure S3 in Supporting Information, which shows the spatial distribution of STS curves crossing the monoiodine vacancy. These experimental results are consistent with the previous theoretical calculations by the Tkatchenko and Scheffler van der Waals correction methods, where the calculated bandgap values of pristine monolayer $BiI_3$ and monoiodine vacancy structure are coincident with each other [36]. The ability to retain the electronic properties of pristine form in the presence of point defect indicates that $BiI_3$ is a monoiodine vacancy-tolerant semiconductor. This defect-tolerant behavior toward anion vacancy is caused by the large mass of Bi intensified spin-orbit coupling (SOC), which broadens the conduction band and pushes CBM to a lower energy than the atomic Bi $6p$ states [9,21].

As stated above, the $BiI_3$ thin film adopt the SK growth mode, leading to abundant edges in the $BiI_3$ thin film sample. Thus, it is necessary to investigate the possible edge effect on the electronic properties of pristine film. The line profile in Figure 4b collected along the white dashed line in Figure 4a shows the spatial distribution of the STS curves around the sample edge. When the measured position is far away from the edge, the gap value is 2.5 eV. And the value rapidly decreases down to 1.5-1.6 eV with both of the VBM and CBM shifting to Fermi level when the point is close to the edge. Therefore, we attribute the bandgap decrement to the effect of the edge state residing in the energy gap. Figure 4c displays the relationship between the gap value and the distance to the edge characterized in Figure 4a, where the effective range of edge state is around 2 nm, which is similar to the size of edge state in previous reports [37-38]. It is worth noting that the gap value near the edge coincides with the results characterized by the optical measurements [3-4,6,8-10,14-15,27-28], inferring that the edge state plays crucial role in these macroscopic measurements. In fact, the photoluminescence spectra indicate a larger energy gap in single crystal compared to the polycrystalline $BiI_3$ [9], which could be ascribed to the edge effect explored in our investigation as the edges are expected to be abundant in polycrystalline samples.

To gain insight into the mechanism of the edge effect on $BiI_3$, the partial charge densities of VBM and CBM, band structures, and orbit-projected DOS of monolayer $BiI_3$ and $BiI_3$ nanoribbon were computed, as shown in Figure 5. For monolayer $BiI_3$, the CBM is mainly contributed by the Bi-$p$ orbits, and the VBM is derived from the $p$ orbits of I atoms, as displayed in Figure 5a, Figure 5b, and Figure S4 in Supporting Information. The band structures of monolayer $BiI_3$ in Figure 5b reveal an indirect energy gap of 2.46 eV by employing the generalized gradient approximation in the

Perdew-Burke-Ernzerhof form (GGA-PBE). All these results are consistent with the previous calculations of the electronic properties of monolayer BiI$_3$ [26,36]. There are two kinds of BiI$_3$ nanoribbon with different species of the edge, where one termination is composed of only I atoms as shown in Figure 5c and the other one is made up by both of I atoms and Bi atoms shown in Figure S5 in Supporting Information. We choose the former as it is with the lower formation energy to make a comparative study to the electronic properties of monolayer BiI$_3$. In this nanoribbon, the edge effect on the electronic structures becomes remarkable, leading to that both of the CBM and VBM are dominated by the $p$ orbits of I atoms, as shown in Figure 5d. Importantly, the value of the energy gap downshifts from 2.46 eV for monolayer to 1.69 eV for the nanoribbon. The deviation value (0.77 eV) is comparable to the difference (~0.6 to 1.0 eV) between the gap value of few-layer BiI$_3$ films measured by our STS curves and the bandgap characterized by macroscopic optical measurements, confirming the edge effect on bandgap decrement of BiI$_3$. We note that the gap value calculated by the PBE method may not be as accurate as the way performed by the Heyd-Scuseria-Ernzerhof functional with consideration of the SOC strengths of Bi and I atoms. Nevertheless, our theoretical results provide a qualitative analysis of the bandgap values of monolayer BiI$_3$ and its nanoribbon, which already shows an excellent accordance with the experimental results.

## 3. Conclusion

In summary, we have successfully grown monolayer and few-layer BiI$_3$ on graphene-terminated 6H-SiC(0001) substrate by molecular beam epitaxy, as identified by the STM results. The weak interaction between adjacent BiI$_3$ layers is revealed by the absence of Moiré pattern in the twisted bilayer BiI$_3$, which results in neglectable dependence of bandgap value on layer thickness. The STS spectra reveal the intrinsic bandgap of 2.8 eV for monolayer BiI$_3$, which could be modulated by the edge state rather than monoiodine vacancies. Utilizing first-principles calculations, we have shown that the I-terminated nanoribbon is responsible for the reduced energy gap of ~1.6 eV observed in the optical measurements in previous reports. Our results provide a general means to accomplish epitaxial growth of atomic thin BiI$_3$ film and to engineer their electronic structures, which is promising for applications in both nanoelectronics and optoelectronics.

## 4. Materials and methods

*4.1. Synthesis of BiI$_3$ on graphene-terminated 6H-SiC(0001) substrate*

All experiments were performed with a low-temperature STM (SNOM 1400, Unisoku Co.). The 6H-SiC(0001) substrate was initially outgassed at ∼900 K, and then flashed up to ∼1600 K for a few cycles until graphene termination was achieved. Then, high quality BiI$_3$ powder were evaporated on the graphene surface under ultra-high vacuum (UHV) conditions from a heated well-degassed Ta wafer. The deposition flux was ~0.02 monolayer per minute (ML/min). The substrate temperature was maintained at room temperature during the deposition process.

*4.2. Characterization of crystalline structure and electronic properties*

All the STM images herein presented were acquired at liquid nitrogen temperature (77 K) with the PtIr tip in constant current mode and the bias voltage applied to the sample. Differential conductance, dI/dV, spectra were acquired by using a standard lock-in technique with a modulation amplitude of 10 mV at a frequency of 973 Hz.

*4.3. DFT calculations*

All calculations were based on the density functional theory, which were implemented in the Vienna Ab initio Simulation Package (VASP) [39-43]. The cutoff energy is set to be 500 eV and the Brillouin zone sampling meshes are set to be dense enough to ensure convergence (11 × 11 × 1 for 2D structures and 1 × 11 × 1 for nanoribbons). The vacuum layers with a thickness of at least 15 Å is used to avoid spurious interactions between adjacent images. All the structures are fully optimized until the residual forces on every atom is less than 0.001 eV/Å.

## Data Availability

The raw/processed data required to reproduce these findings are available from the 484 corresponding author on reasonable request.

## Declaration of competing interest

The authors declare no conflict of interest.

## Acknowledgements

D.M. and W.Z. contributed equally to this work. The work was supported by the Beijing Natural Science Foundation (Z180007), the National Natural Science Foundation of China (11874003,

## Supplementary Material

Supplementary Matrial to this article can be found online.

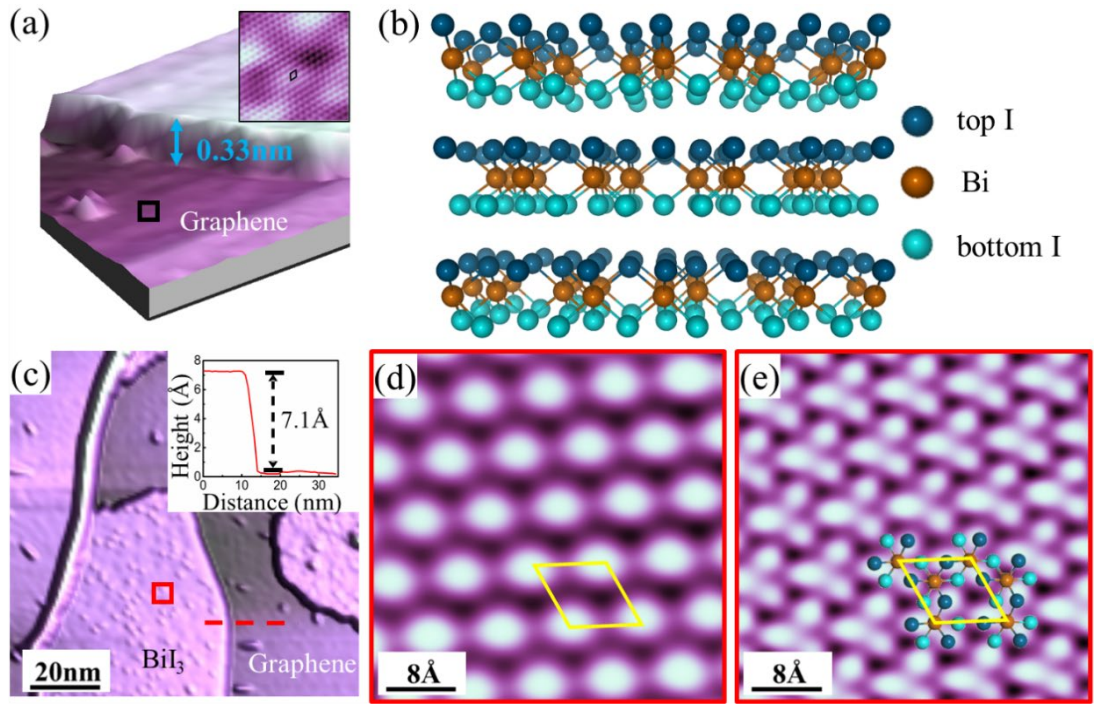

Fig 1 a) STM topographic image of graphene terminated 6H-SiC(0001) ($V_{bias}$ = 2.0 V, $I$ = 50 pA, 50 nm × 50 nm). Inset shows the high-resolution STM image of graphene in the region label by black square in panel a ($V_{bias}$ = 300mV, $I$ = 50 pA, 4 nm × 4 nm). The black rhombus stands for the unit cell of graphene. b) Atomic structure of $BiI_3$. The blue, yellow, and green balls denote the top I atoms, Bi atoms, and bottom I atoms in one $BiI_3$ layer, respectively. c) Differential STM image of sub-monolayer $BiI_3$ film on graphene ($V_{bias}$ = 2.0 V, $I$ = 50 pA, 100 nm × 100 nm). Inset: line profile corresponding to the red dashed line in panel c. d) High-resolution ($V_{bias}$ = 1.5 V, $I$ = 100 pA, 4 nm × 4 nm) and e) atomic resolution ($V_{bias}$ = 1.0 V, $I$ = 100 pA, 4 nm × 4 nm) STM images of $BiI_3$ measured in the same area labeled by the red square in panel c. The yellow rhombus represents the unit cell of the $BiI_3$. "L" stands for the layer.

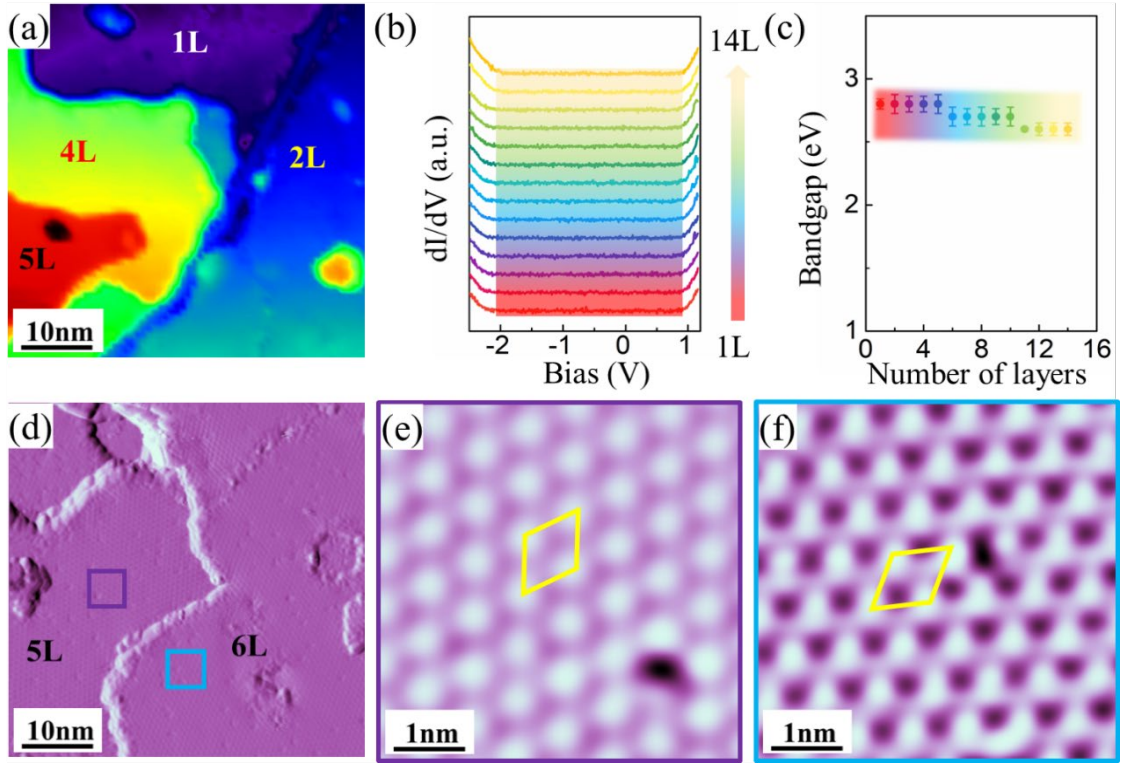

Fig 2 a) Morphology of multilayer BiI$_3$ film ($V_{bias}$ = 2.0 V, $I$ = 50 pA, 50 nm × 50 nm). b) STS curves measured in different layers (set point: $V_{bias}$ = 2 V, $I$ = 300 pA). c) Derived bandgap values as a function of layer numbers from panel b. d) Differential STM image of multilayer BiI$_3$ ($V_{bias}$ = 1.5 V, $I$ = 100 pA, 50 nm × 50 nm). High-resolution STM images of e) 5 L ($V_{bias}$ = 1.5 V, $I$ = 300 pA, 5 nm × 5 nm) and f) 6 L BiI$_3$ ($V_{bias}$ = 1.3 V, $I$ = 300 pA, 5 nm × 5 nm) corresponding to purple square and blue square in panel c, respectively.

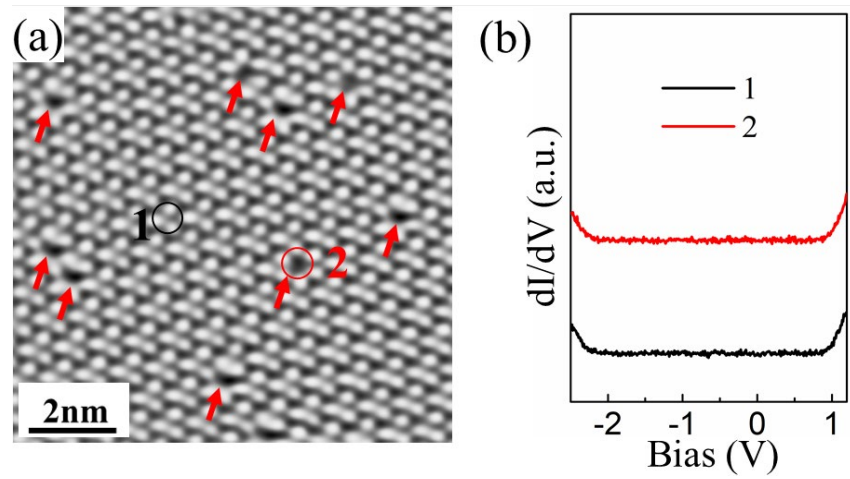

Fig 3 a) STM image of BiI$_3$ film with monoiodine vacancies marked by the red arrows ($V_{bias}$ = 1.0 V, $I$ = 100 pA, 10 nm × 10 nm). b) STS spectra of pristine BiI$_3$ (labelled as point 1) and monoiodine vacancy (labelled as point 2) in panel a, respectively (set point: $V_{bias}$ = 2 V, $I$ = 300 pA).

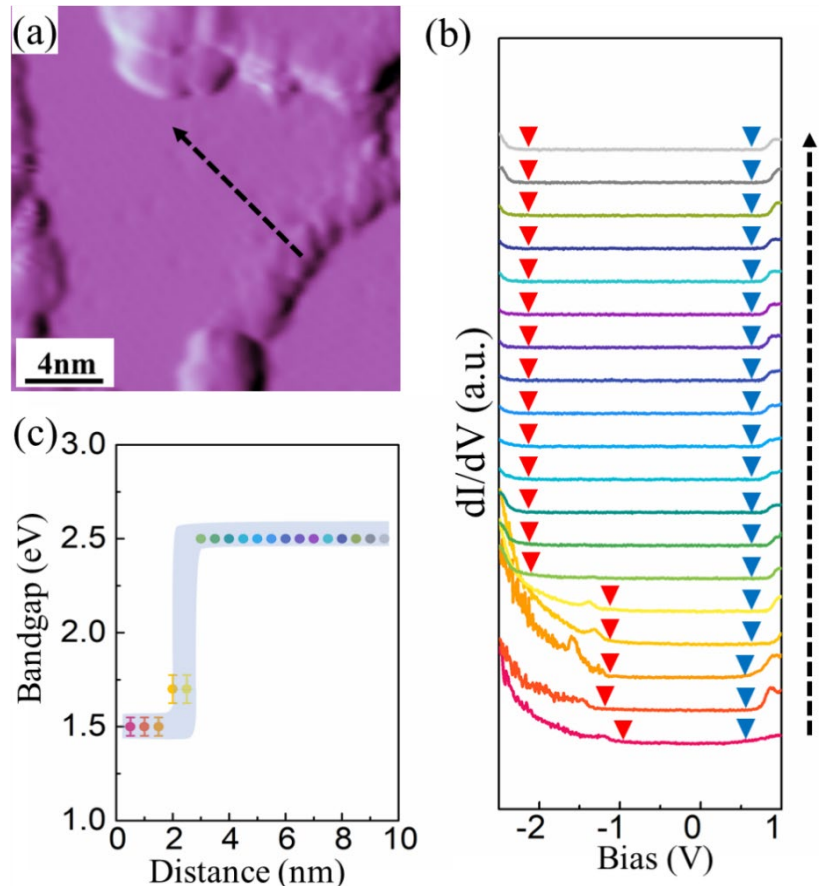

Fig 4 a) Differential STM image of multilayer BiI$_3$ ($V_{bias}$ = 2.0 V, $I$ = 50 pA, 20 nm × 20 nm).

b) Tunneling conductance curves along the dotted arrow in panel a (set point: $V_{bias}$ = 1 V, $I$ = 300 pA). The red and blue inverted triangles represent the energy positions of VBM and CBM, respectively.

c) Derived bandgap values as a function of the distance to the edge from the STS curves in panel b.

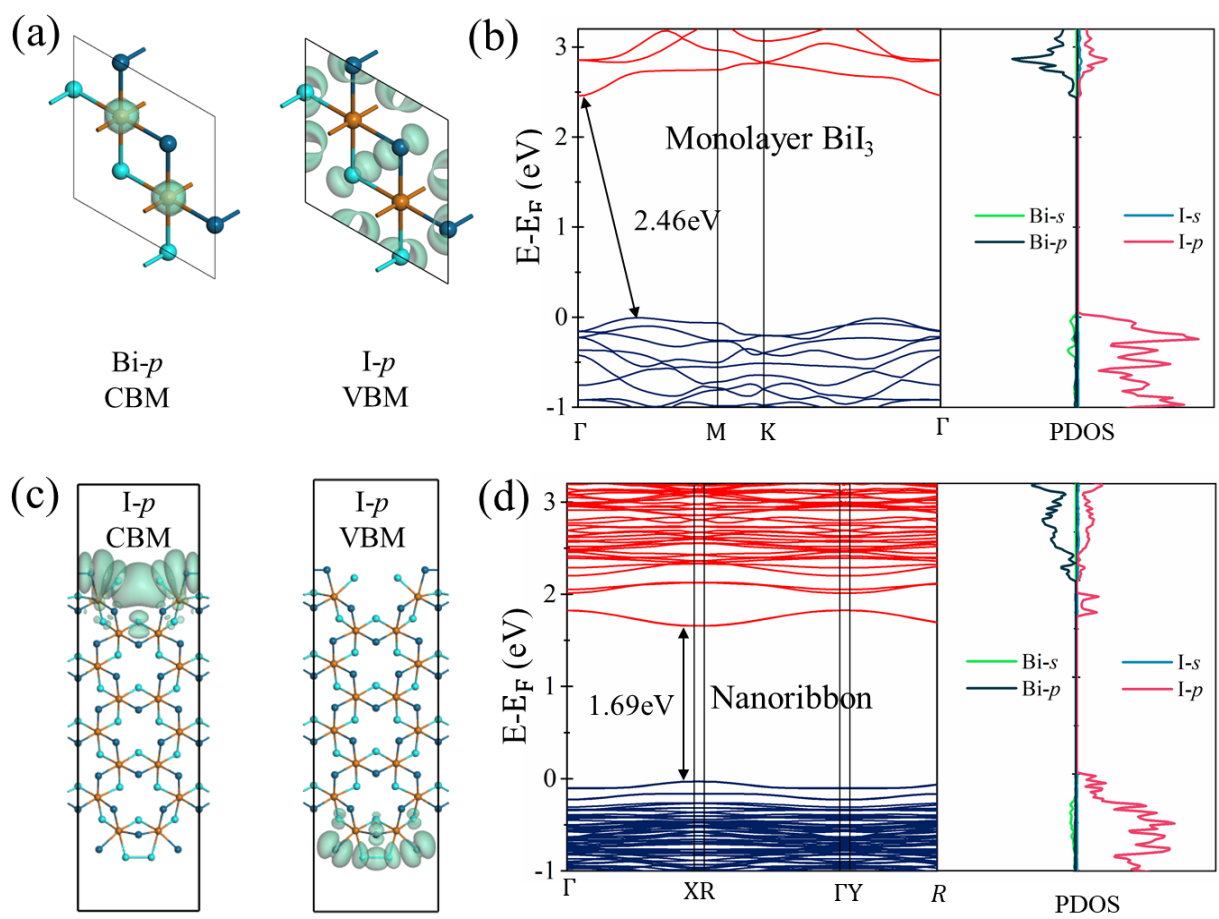

Figure 5 Partial charge densities of CBM and VBM of a) monolayer $BiI_3$ and c) $BiI_3$ nanoribbon, respectively. b) Calculated electronic band structure and the orbit-projected DOS of b) monolayer $BiI_3$ and d) $BiI_3$ nanoribbon using PBE method, respectively.